\documentclass[11pt]{article}
\usepackage[T1]{fontenc}
\usepackage[utf8]{inputenc}
\usepackage{mathptmx}
\usepackage[protrusion=true,expansion=true]{microtype}
\usepackage[margin=1in]{geometry}
\usepackage{amsmath,amssymb}
\usepackage{graphicx}
\usepackage{placeins}
\usepackage{enumitem}
\usepackage[round,authoryear]{natbib}
\usepackage{xurl}
\usepackage[hidelinks]{hyperref}
\hypersetup{pdftitle={After One-Click Video: The Inflow of Service-Type Content, the Failure of the Recommendation Pool Mechanism, and the BoLe System},pdfauthor={Bo Zou and Chao Xu}}
\setlist{nosep,leftmargin=*}
\DeclareUnicodeCharacter{00D7}{\ensuremath{\times}}
\providecommand{\tightlist}{\setlength{\itemsep}{0pt}\setlength{\parskip}{0pt}}
\begin{document}
\begin{center}
{\LARGE\bfseries After One-Click Video:\par}
\vspace{0.3em}
{\Large The Inflow of Service-Type Content, the Failure of the Recommendation Pool Mechanism, and the BoLe System\par}
\vspace{0.8em}
{\large Bo Zou \qquad Chao Xu\par}
\end{center}
\begin{abstract}
In September 2026, a new large language model was released. Soon after, a new way of making explainer videos became very popular: people asked the model to write a program, and the program produced the video. ``One-click video'' content then appeared in large amounts on many platforms. On the surface, this is a rise in creative power at the level of the individual. We argue that, at a deeper level, it changes the structure of the content industry. This judgment rests on two observations. First, this wave of content is service-type content. People watch it to meet a need, not to enjoy the content itself. The other kind is work-type content. AI makes both kinds faster to produce. But anyone can come in and make service-type content, while work-type content is limited by the number of people who have something to say. So the ratio between the two slowly becomes unbalanced. Second, free platforms choose content through layers of trial viewing, and this uses up user attention. The total amount of attention is fixed. The unbalanced ratio also dilutes good content. A simplified mathematical model shows that, as long as these conditions hold, the recommendation pool mechanism will fail once new content grows past a certain multiple. Based on this, we discuss one possible way out. Machines and entry rules judge whether service-type content passes the bar. Trusted people judge work-type content. We call this the BoLe system. At the end, we briefly imagine how the two kinds of content may develop from here.
\end{abstract}
\section{Introduction}

\subsection{What happened}

Using machines to mass-produce content in exchange for traffic and ad revenue share is not new. More than ten years ago, content farms had already turned ``choose topics by ad price, write cheaply from templates'' into the business of a listed company \citep{roth2009}.

In autumn 2026, a new turning point came. Opus 5.5 was released on September 22. After that, a new method spread fast on social platforms: use the model to write a program that draws each frame and puts the frames together into a motion graphics video. These videos do not go through a video generation model. Instead, the model writes a program that draws every frame, and the program is then rendered into a video file \citep{dsebastien2026}. One week after the release, the industry newsletter Latent Space wrote in a review that these code-generated explainer videos had ``taken over the timeline'' \citep{latentspace2026}. Once the method was understood, other top-tier models could also produce similar results. Soon, ``one-click video'' skill packs and tutorials were everywhere on all platforms. The platforms have not published numbers for this kind of content. The scale of the wave is the authors' observation as people who work in the field.

The first author has direct experience of how fast this is. Our team has made professional short videos for more than ten years. In the past, a five-minute science video built mainly on motion graphics took five days. We first made the effects frame by frame in After Effects, then edited and put it together in Final Cut Pro. Now, with the first author checking each video in person and taking responsibility for the content, we can make three to five videos in one day.

Readers who study AI systems may not follow these changes in the content market. So we first explain them clearly. The next question is this: so many beautiful videos are pouring out, but what exactly are they?

\subsection{Both are content, but not the same product}

The large number of beautiful videos in this wave is, in essence, a service, not content in the traditional sense. Their goal is to serve. People watch them to meet a certain need, not to ``enjoy'' the work itself.

\begin{quote}
\textbf{Service-type content}: Its goal is to meet a certain need of the viewer. People watch it for that need. Example: most AI short dramas.

\textbf{Work-type content}: People watch it to enjoy the content itself. Examples: a long video in which a big YouTube creator explains some event; a Netflix series; a Stephen King novel.
\end{quote}

The difference between the two is not production cost, and it is not length. A one-click video and a long video by a big YouTube creator can sit in the same feed. They can play in the same player. Both can live on ad revenue share. Even if we make the two the same length and spend the same money on them, viewers still get two different things from them.

Seeing these two kinds of content clearly helps several groups: the producers who are now pouring in, the platforms that run content, and the people who build AI creative tools and models. Which kind they are dealing with decides what they should value and how they should measure success. But the bigger question right now is this. Production efficiency has jumped by one to two orders of magnitude within a few weeks. Much more service-type content than work-type content is pouring in. What will happen to the current content ecosystem?

\subsection{Why this matters}

At first glance, one-click video raises the efficiency of creators. That is true. Every person now has a production team. Individual creative power has grown a lot. It looks like a creative revolution at the level of the individual. But at a deeper level, it changes the structure of the whole content industry.

How can a rise in individual ability become a change in the structure of an industry? The restaurant business helps us understand this. In the past, running a restaurant depended on cooking skill. A skilled cook made dishes. After the food industry matured, semi-finished food and pre-made ingredients let an ordinary person with no cooking skill open a shop. If he opens a breakfast stall, what he provides is a service: food that fills you up, is clean, fast and cheap. It is not cooking skill. For this person, his ability has grown a lot. For the whole restaurant business, there are now many more breakfast stall businesses, and the makeup of the industry has changed. The one-click videos in this wave are the breakfast stalls of the content business.

Our judgment rests on two observations. First, AI makes both kinds of content faster to produce, but service-type content can grow much more than work-type content, so the ratio between the two slowly becomes unbalanced. Second, free platforms choose content through user trial viewing. When content grows very fast and good content is diluted, this method will fail. The first observation is one of the causes of the second.

\subsection{Scope}

This paper discusses free content platforms that distribute content through feeds. Content is free to watch, the platform earns money from ads, and producers get a share. Subscription services, or pay-per-item services where viewers come for ``works'', are outside the scope of our argument.

\subsection{Contributions}

\begin{enumerate}
\def\labelenumi{\arabic{enumi}.}
\tightlist
\item
  We propose the difference between service-type content and work-type content, and explain how AI affects the supply of the two kinds in different ways.
\item
  We argue that when content grows very fast, the recommendation pool mechanism of free platforms, which relies on trial viewing with user attention, will fail. We show this with a simplified mathematical model.
\item
  Based on this, we propose the BoLe system as one possible way out.
\end{enumerate}

The rest of the paper is organized as follows. Section 2 analyzes how AI affects the supply of the two kinds of content in different ways. Section 3 builds a simplified mathematical model and argues that the current recommendation pool mechanism will fail when supply rises sharply. Section 4 discusses the BoLe system as one possible way out. Section 5 imagines how the two kinds of content may develop and what this means for producers. Section 6 discusses limitations, and Section 7 concludes.

\section{The first observation: the ratio between the two kinds becomes unbalanced}

Using AI to write code and make videos has become very easy. Which kind of content benefits? Both kinds. AI can do one part of the job for people: doing the work well by following clear requirements. Authors of work-type content also become much faster because of this. The first author's own example in Section 1.1 shows it.

But the two kinds will not grow by the same amount, because their upper limits for growth are different. For service-type content, anyone can come in and make it, as long as the work is done properly. If you do not check quality and only want the video to look good, you can go even faster. Work-type content needs creators who want to express something, and the number of such people will not multiply just because tools get better. At the time of writing, AI still cannot decide for the producer what to say. So over time, service-type content will take a larger and larger share of all content.

This kind of thing has happened many times in human history. Photography is the example closest to daily life. In the mid-nineteenth century, to take a photo you had to coat a glass plate with chemicals on site, expose it while it was still wet, and develop it right away. A photographer had to carry a darkroom. Photography was a craft of the photo studio. In 1888, Kodak launched a box camera loaded with film for one hundred photos. Its slogan was ``You press the button, we do the rest.'' After you finished shooting, you mailed the whole camera back to the factory. The factory developed the film, loaded new film, and mailed the camera back. In 1900, the Brownie camera, at one dollar each, let even children take photos. From then on, snapshots of family groups, birthdays and trips became part of daily life in millions of homes. Taking photos no longer needed craft. It was only after this that the work-type side got a name of its own. In 1902, Alfred Stieglitz and others started the Photo-Secession. They called themselves art photographers and set themselves apart from the snapshots everywhere. The clothing industry had a similar experience. In the age when ready-to-wear clothing spread, haute couture got its own trade association and a name protected by law.

Code-generated video has also made video production extremely easy. The sharp rise of service-type content, as it has happened many times in history, is now happening in video. Those affected are viewers, creators, video platforms, and the advertisers behind them.

\section{The second observation: the recommendation pool mechanism will fail}

\subsection{How platforms live, and how they choose}

This section discusses video platforms. They have one important business model: users watch for free, the platform earns money from ads, and producers get a small share. For example, on YouTube long videos, creators get 55\% of the net ad revenue (YouTube official). For one of the first author's popular WeChat articles, the revenue share was from a dozen to a hundred or more yuan, and the platform's income was many times what the first author got.

There is so much new content every day. How does a platform decide which item to push? It relies on what people in the industry call the recommendation pool mechanism, also called the ``traffic pool''. Take an unknown new creator as an example. His new item is first pushed to about one hundred people who may be interested, as trial viewers. If it performs above the average of similar items, it advances to a pool of about five hundred people, and then it is amplified layer by layer. If it fails at any stage, it stops at that layer (the numbers are only for illustration). Platforms use this method because user feedback is the most accurate and also the cheapest kind of judgment. A platform cannot hire tens of thousands of editors. In recommender system research, this matches the cold-start problem for new content: how to allocate the first exposure of new content between ``exploration and exploitation'' \citep{li2010,wu2016}.

\subsection{Trial viewing costs attention}

This method has a premise that is rarely said out loud. Trial viewing costs user attention, and the total amount of attention is fixed. Simon pointed out half a century ago that a wealth of information creates a poverty of attention \citep{simon1971}.

Each trial view trades a few seconds of one user's time for a little information about whether this item is good. Suppose one trial view brings back on average \(c\) bits of information about quality. Suppose the number of trial views the platform can use each day is \(\varepsilon X\), and the number of new items is \(N\). Then the total amount of such information the platform can collect each day is at most \(c\,\varepsilon X\) bits. Each item gets on average

\[I_{\text{per item}}\le\frac{c\,\varepsilon X}{N}.\]

The numerator is set by users' time. It does not grow when content grows. If \(N\) grows one hundred times, each item gets one hundred times less information. To judge whether an item is good, you need to collect a certain amount of information. If an item gets less than that amount, the judgment is close to a guess. In the past, content was slow to make and new content was scarce. Trial viewing took only a small part of users' time, and the platform could afford it. The machine learning literature has studied the problem of ``more options than the number of trials you can run'' \citep{wang2008}. It treats this as an algorithm problem and asks how to run trials more cleverly. This paper asks who pays this cost, and whether they can afford it.

\subsection{Neither path works}

When new content grows to \(k\) times its original amount, the platform has only two paths.

Path one: the total trial-viewing budget stays the same, and the number of trial viewers per item falls to \(1/k\) of the original. The fewer people there are, the more the observed completion rate is driven by random noise. Good and bad can no longer be told apart, and selection becomes more and more like drawing lots.

Path two: the number of trial viewers per item stays the same, and the share of users' time spent on untested content rises to \(k\) times the original. If the baseline is 10\%, then at ten times, everything users see is untested content.

The platform can only find a compromise between the two paths, and both ends are bad.

If good content also grows a little, can that make up for it? No.~In this section, the unbalanced ratio from Section 2 becomes the dilution of good content. The share of good content among the items that advance is given by Bayes' formula:

\[P(\text{good}\mid\text{advance})=\frac{P(\text{advance}\mid\text{good})\,P(\text{good})}{P(\text{advance}\mid\text{good})\,P(\text{good})+P(\text{advance}\mid\text{mediocre})\,P(\text{mediocre})}.\]

Here is a set of illustrative numbers. Suppose trial viewing is quite accurate: 90\% of good content advances, and only 10\% of mediocre content advances. When good content is 5\% of new content, about 30\% of the items that advance are good. When good content is diluted to 0.05\%, less than 0.5\% of the items that advance are good. Trial viewing itself has not changed at all, but almost everything that advances is mediocre content that got lucky. Once the base rate falls, even very accurate trial viewing cannot save it.

We use a toy model with three layers of trial viewing to show this (Figure 1; this is a simplified mathematical model, all parameters are illustrative values, and the code is public). In the baseline, good content is 5\% of new content. The number of trial viewers per item in the three pool layers is 100, 500 and 2500, and the total trial-viewing budget of each layer is fixed. The result is as follows. While new content grows up to three times, the mechanism still has room, and the top pool is almost all good content. Past this point, the share drops quickly. At ten times, good content in the top pool is only about 12\%. At one hundred times, it is only about 0.1\%. Even if we assume that good content also grows by the same ratio, which is the most optimistic case, at one hundred times good content in the top pool is only about 10\%. This is only twice the 5\% at input, so selection almost does nothing.

In the toy model, the growth multiple of new content is a parameter. We take it up to one hundred times based on the first author's own experience. When we drop professional quality checks and only want the video to look good, we can make twenty videos a day, about one hundred times the past. People who turn to mass production will not check every item. If the parameters take other values, the direction of the conclusion does not change. As long as two things hold, ``trial viewing costs user attention'' and ``the total amount of attention is fixed'', selection will fail once new content grows past some multiple. The parameters only decide where that multiple falls. We tested this by setting each parameter to half and to double its value. In all cases, the higher the multiple, the less good content in the top pool. When we change one parameter at a time, the multiple at which the share of good content falls below half is between about three and forty times. We also tried the case most favorable to the platform. We doubled three parameters at the same time: the share of good content, the gap in completion rate between good and bad content, and the number of trial viewers per layer. Even then, this multiple is only pushed back to between fifty and nearly two hundred times. And this combination means users would have to spend 20\% of their time on trial viewing.

\begin{figure}[t]
\centering
\includegraphics[width=\textwidth]{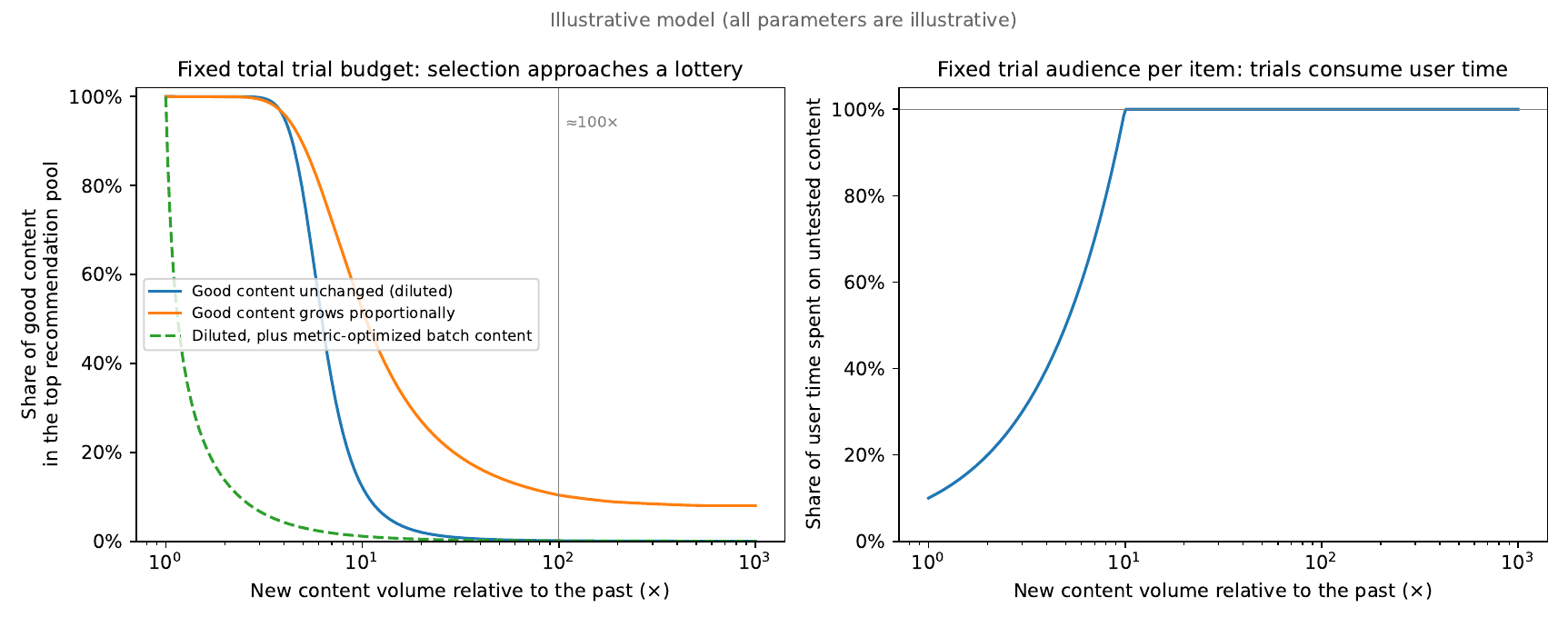}
\caption{The dilemma in the toy model. Left: with a fixed total trial-viewing budget, the share of good content in the top layer of the recommendation pool as new content grows by different multiples. The three lines are: the amount of good content stays the same; good content grows by the same ratio; the added mass-produced content is optimized for the metrics. Right: with a fixed number of trial viewers per item, the share of users' time spent on untested content. The horizontal axis is on a log scale, and all parameters are illustrative values.}
\label{fig:dilemma}
\end{figure}

\subsection{Content made to target the algorithm}

There is another type of producer. They make content to earn traffic and revenue share, not to express something. They do not wait passively to be chosen. Instead, they study what the recommendation algorithm values and make content that targets the trial-viewing metrics.

Every item has two things. One is the first taste at the entrance: the cover and the first three seconds. Trial viewing can measure this. The other is the substance you swallow: how good it really is. There is a well-known question: if you had to eat one, would you choose poop-flavored chocolate or chocolate-flavored poop? The trial-viewing mechanism gets both of them wrong. Chocolate-flavored poop has good metrics and bad substance, and it is let into the big pools. Poop-flavored chocolate has good substance but is not pleasant at the entrance, and it is kept out.

In the language of information theory, optimizing for the metric lowers the mutual information \(I(Q;M)\) between the metric \(M\) and the true quality \(Q\). No matter how much you measure the metric, it brings back less and less information about true quality. This problem has nothing to do with the trial-viewing budget. In the toy model, if all the added mass-produced content just passes the bar on the trial-viewing metrics, then when new content grows only three times, good content in the top pool is only about 7\%.

Platform subsidies make this worse. During cold start, platforms often give subsidies based on quantity or metrics. A producer's margin equals the subsidy minus the production cost. When production cost is pushed toward zero, the margin gets close to the full subsidy. This attracts more mass producers.

History has seen a rehearsal of this. Demand Media used algorithms to find search terms with high ad prices. Freelance writers got 15 to 20 dollars per piece, and the company produced more than four thousand articles or videos every day \citep{roth2009}. In 2011, Google's ``Panda'' update affected about 11.8\% of search queries in the United States. It was aimed at this kind of low-quality site \citep{singhal2011}.

\subsection{Letting a machine review first is not enough}

A natural idea is to let AI review content first and block the bad items. This is not enough, because the platform does not only want to pick out good content. It has to take care of a whole set of needs at the same time:

\begin{enumerate}
\def\labelenumi{\arabic{enumi}.}
\tightlist
\item
  Find good content and show it to people;
\item
  Keep the creators who can attract advertisers, so they do not leave for other platforms;
\item
  Not be filled with large amounts of content nobody watches, because storage, distribution and moderation all cost money;
\item
  Remove producers who mass-produce content to cheat traffic;
\item
  Keep users, and let them scroll comfortably;
\item
  Keep advertisers satisfied;
\item
  Every item must pass compliance review. If content grows one hundred times, review work also grows one hundred times.
\end{enumerate}

Machine pre-review can at most take care of parts of needs 1, 3 and 4, and it also costs money per item. It cannot take care of need 2. Whether creators leave depends on whether they can be seen and whether they can earn money, and the revenue share has already been thinned out by much more content (see Section 3.7). Machine pre-review can tell whether an item passes the bar. It cannot tell whether the item expresses something well. It can block the garbage below the bar. It cannot block the large amount of service-type content above the bar.

When content grows very fast, these needs also fight each other. To find the next new talent, the entrance must be opened wider. To block mass production, the entrance must be tightened. In the past, most newcomers were serious, and the two sides did not conflict. Now the two sides crash into each other directly. When the platform makes trade-offs in this conflict, it follows a utilitarian rule. It can accept that a creator who can attract advertisers gets buried, but it cannot accept that he runs off to another platform. So when content floods in, the platform first protects the old accounts that have already proved they can attract ads, and the entrance for finding new people narrows. More and more people will be buried. The platform can accept this. New people have to find other ways to be seen.

In reality, some platforms are already tightening the entrance. A paid science fiction magazine blocked more than five hundred generated spam submissions in the first twenty days of February 2023. The editors could not review them all, so they closed submissions \citep{clarke2023}. In September of the same year, Amazon's self-publishing platform set a rule that each account can create at most three new books per day \citep{guardian2023}. A large programmer Q\&A community banned answers written by generative tools in December 2022. One of the reasons was that the moderators were overwhelmed \citep{stackoverflow2022}.

\subsection{The production barrier used to do the first round of selection for the platform}

So many problems appear at the same time, and they have one cause. Before code-generated video, making content was hard. Few people were willing to make it and able to make it. The production barrier kept most people out. In effect, it did the first round of selection for the platform for free. Each of the needs above quietly depended on this. Now the barrier has dropped, this round of selection is gone, and all the needs fall out of balance together. Section 2 said that AI makes service-type content grow. What we see here is the other side of the same thing.

\subsection{The revenue share is thinned out}

For producers who live on revenue share, this means the same pool of ad revenue is split among much more content. Each item gets less and less. In 2020, TikTok set up a fixed 200 million dollar creator fund in the United States, split by share of views. After the number of creators rose sharply, the fund was thinned out. One well-known creator reported that his pay fell from about 5 cents to 2.5 cents per thousand views, because the same fixed pool was shared by more and more creators \citep{hale2022}. The fund was shut down in December 2023 \citep{nbc2023}. Even before this, on YouTube the top 3\% of channels took more than 85\% of views, and about 96.5\% of creators who wanted to earn money from ads could not reach the US poverty line \citep{bartl2018}.

Taking this whole section together, we believe the current recommendation pool mechanism will fail. The root cause is that the platform has only one source of information for judging content quality: user attention. And its total amount is fixed. No matter how the platform adjusts things inside, it cannot create more attention. So, to judge content quality, another source is needed, and this source must not spend user attention again.

To make the loop in this section visible, we built a demo with light interaction (Figure 2). The viewer adjusts only one parameter, the production multiple brought by AI, and then watches a free platform run for one year. The selection part of the demo uses the same algorithm and parameters as Figure 1. User churn, ad revenue, revenue share, platform costs and creator churn are illustrative rules added on top of that. When production is at the past level, the platform stays stable for the whole year. When the multiple goes above three to five times, the recommendation pool can no longer pick out good content. Users and ad revenue then decline, while the platform's costs rise with the amount of content.

\begin{figure}[!htbp]
\centering
\includegraphics[width=0.92\textwidth]{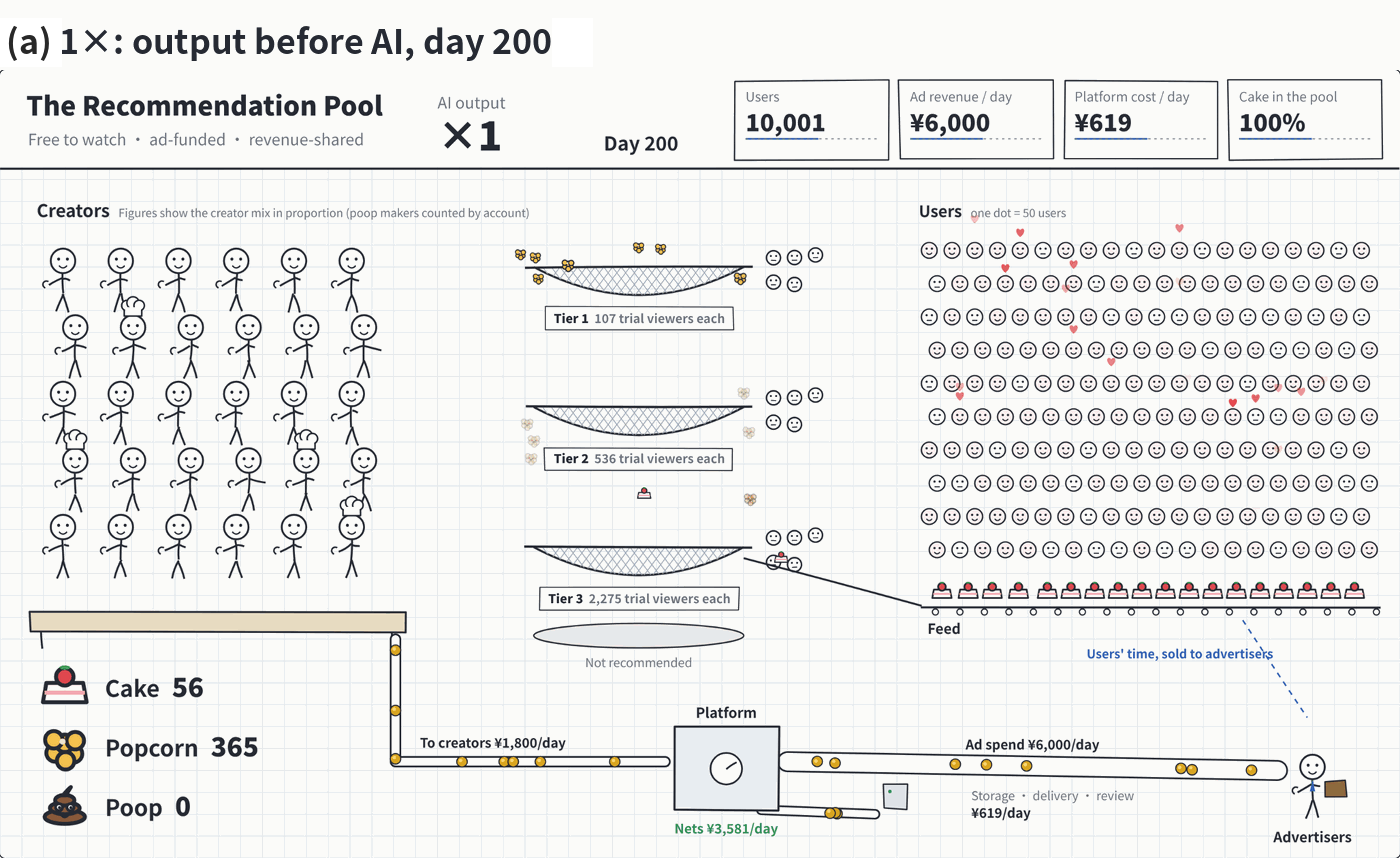}\\[0.6em]
\includegraphics[width=0.92\textwidth]{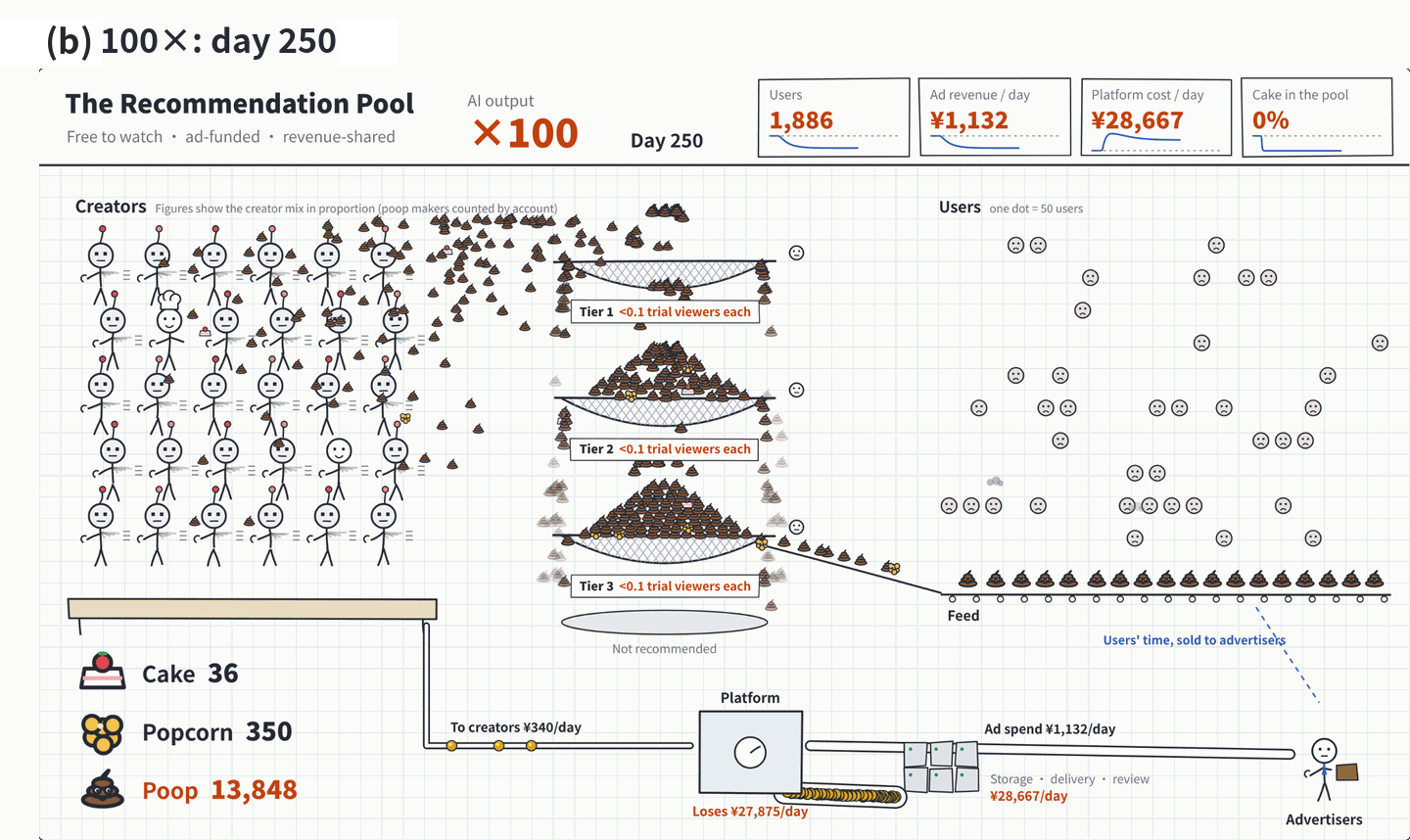}
\caption{Two moments in the recommendation pool demo. (a) Day 200, with production at the past level. (b) Day 250, with production at one hundred times the past level. In the figure, things flow from left to right, and money flows back from right to left. Creators put out content, three layers of sieves run trial viewing, and users eat the results in the recommendation feed. Advertisers pay for users' time. The platform keeps the larger part, gives a small part to creators, and also pays the costs of storage, distribution and moderation. Cake, popcorn and poop stand for good content, ordinary content and mass-produced content. The selection part uses the parameters of Figure 1. The rules and parameters of the loop part are illustrative and have not been calibrated with real data.}
\label{fig:demo}
\end{figure}

\FloatBarrier

\section{One possible way out: the BoLe system}

\subsection{The two kinds of content need different kinds of judgment}

Can machines be this source? For service-type content, yes. Service-type content exists to meet a need. Once the need is met, that is enough. So it only needs to answer whether it passes the bar, and machines and entry rules can do that. Work-type content is enjoyed for the content itself. Better is always better, and there is no line where it is ``enough''. To judge work-type content, you have to answer how good it is, and this needs much more information. In the language of information theory, passing the bar or not is a one-bit question. If you want to sort work-type content into \(K\) levels, you need about \(\log_2 K\) bits. Machines can give the former. The latter needs a source with much more information.

\subsection{The BoLe system}

People who know the field can give this kind of information. When an expert takes a careful look, they bring back much more information than an ordinary viewer's quick swipe. And their judgment does not use the attention of the general public. In the terms of Section 4.1, one swipe by an ordinary viewer brings back only a little information about quality. It is also mixed with a lot of random factors that have nothing to do with the content: their finger may have slipped, or they may just not have been in the mood at that moment. When an expert watches carefully once, they can say what is good and what is bad. The information they bring back may be worth hundreds or thousands of swipes, with much less noise. This is a channel with large capacity and low noise. It fills in exactly the bits that work-type content needs.

Bo Le was a man in ancient China known for judging horses. Later, his name came to mean anyone good at spotting talent. We call this kind of arrangement the BoLe system. Machines and entry rules handle whether service-type content passes the bar. Trusted people judge work-type content. Then machines amplify and distribute their choices.

Early forms of this already exist. In 2023, the American writing platform Medium launched a ``Boost'' mechanism. It works by community nomination, human confirmation and machine distribution. Editors of publications in different fields are invited as nominators, the platform team makes the final decision, and approved stories are then amplified by the algorithm \citep{stubblebine2023}. In 2015, Apple Music set up a team of expert editors to make playlists \citep{apple2015}. In October 2026, the first author of this paper was invited by the WeChat Official Accounts platform to take part in a closed test as a BoLe for ``Original Picks'' (also called ``recommend''). The first author was one of the earlier people invited.

It is interesting that the last time the internet was flooded with information, the direction was the opposite. In the 1990s, the number of web pages exploded. Human catalogs such as the Yahoo Directory could not keep up and were replaced by algorithmic search. This time, algorithmic trial viewing cannot pay the bill, and people are being invited back. The way they come back is also different. People do not look at everything one by one. People choose, and machines amplify. In history, after a flood of goods, trust has often moved from ``try it yourself'' to trusted third parties. Content is a typical experience good: you only know whether it is good after you use it \citep{nelson1970}. The Michelin Guide and the book review sections of newspapers are both third parties that try things first for others, and they appeared after large numbers of similar goods came out.

\subsection{Trust can only be built over time}

If so, can a platform build a BoLe system at any time? It can build the structure, but trust cannot be built fast. A reader's trust in a BoLe is built up one recommendation at a time. Let \(\theta\) be the probability, in the reader's view, that ``what this BoLe recommends really is good''. Let the starting belief be \(\mathrm{Beta}(\alpha,\beta)\). After the reader sees \(m\) recommendations and is satisfied with \(h\) of them, the belief is updated to

\[\theta\sim\mathrm{Beta}(\alpha+h,\ \beta+m-h).\]

Each recommendation changes trust only a tiny bit. So trust can only be built up over time. Money cannot buy it all at once. Michelin took decades, with anonymous inspectors who paid for their own meals.

From this we claim a conditional cumulative advantage. The platform that builds a BoLe system first will gather trusted people first. Following the trade-off rule in Section 3.5, creators who have something to express and can also attract ads will get buried in the flood, and they will move to places where a BoLe can see them. This paper does not judge the rise or fall of any specific platform, because rise and fall is the result of many factors. For example, it also depends on network effects, user habits, the power of capital, and so on.

\section{Possible developments and what they mean for producers}

\subsection{If the two systems of selection take shape}

If the two systems of selection, one by machines and one by BoLe curators, take shape, the two kinds of content may each find a different way to live. Service-type content may turn to clients who pay directly, for example making videos for businesses or doing custom work. Or it may become a stable, cheap base supply on platforms, like breakfast stalls, competing on stability, speed and price. Work-type content may be seen through trusted people, and survive on people who pay for ``this one piece''. Judging itself may also become a business.

These are only ideas. Whatever the future looks like, people who make content now can already do one thing.

\subsection{For producers: first know which kind you make}

This point does not depend on how the future changes. It depends only on two things already shown above: the difference between the two kinds of content, and the failure of the current recommendation pool mechanism.

Getting your position wrong will cost more than before. If you make service-type content in the way you make work-type content, you will put effort into places viewers do not pay for. You will neglect stability and update rhythm, and you will not get your investment back. If you make work-type content in the way you make service-type content, using data in place of judgment and output in place of expression, you will be treated as something that can be swapped out.

Getting the position wrong often does not come from unclear thinking. It comes from an urge that is hard to hold back. An expert works in a field they know and love. They cannot stop themselves from expressing something, and this urge leaks into the production of service-type content. So it is not enough to think it through. You need to run the two kinds separately in the structure itself. Split off the production of service-type content on its own, and make a clear rule that there is no expression here. One team can make both kinds of content at the same time, as long as each side keeps to its own rules.

The experience of our team can serve as a small example. The team used to have one content pipeline that did two jobs: quickly growing new accounts and carefully polishing work. After we split it into two, the pipeline for service-type content cut away all the care spent on expression and polish. What remained were only the steps needed to pass the bar and stay stable. This is only one sample, and it is a description after the fact. It is given only as an illustration.

\section{Discussion and limitations}

The toy model in this paper is a simplified mathematical model. Its parameters are illustrative values. The conclusion depends on the structure, not on the specific numbers. The description of platform trial-viewing mechanisms comes from the authors' work experience, not from material made public by platforms. The judgment that code-generated video spread across platforms in autumn 2026 comes from industry media reports and the authors' observation. The platforms have not published data. The case of our team is only one sample. The demo in Figure 2 adds, on top of selection, a loop of user, ad and creator churn. These rules are illustrative, and the specific numbers in the figure are not used as evidence. This paper offers a judgment and an outlook, and the outlook may be wrong.

The relation to existing theory is briefly as follows. Bourdieu separates the field of restricted production from the field of large-scale production. He discusses whether the standards of a field are set by peers or by the market \citep{bourdieu1992}. Our distinction is made at the level of the single item of content. The ``culture industry'' of Adorno and Horkheimer carries a moral critique \citep{adorno1944}. This paper does not judge which kind of content has more value. It only discusses the position of the two kinds of content in the ecosystem. Certification economics points out that a monopoly certifier tends to tell buyers only ``pass or fail'', and only competing certifiers give detailed quality information \citep{lizzeri1999}. This suggests that BoLe systems may also be monopolistic or competitive.

Finally, we leave one question for future research. When readers follow a BoLe, do they follow him because he chooses well, or because of the person himself? If it is the former, the BoLe provides a service that meets a need. If it is the latter, the BoLe's judgment itself has a work-type side.

\section{Conclusion}

The one-click videos in this wave are service-type content. They look like a rise in individual efficiency, but at a deeper level they change the structure of the content industry. AI makes both kinds of content faster to produce. But anyone can come in and make service-type content, while work-type content is limited by people who have something to say. So the ratio between the two slowly becomes unbalanced. The imbalance dilutes good content. Together with the fixed total amount of user attention, this means the recommendation pool mechanism of free platforms, with its layers of trial viewing, will fail. As one possible way out, this paper discussed two systems of selection: machines handle whether service-type content passes the bar, and trusted people judge work-type content. This is the BoLe system. For people who make content now, the most important thing is to first know which kind they make.

\section*{AI Use Statement}
\addcontentsline{toc}{section}{AI Use Statement}

AI tools were used for the discussion, literature review, simulation code and first draft of this paper. The discussion and thinking were done mainly with the help of Opus 5.5, which is also one of the subjects discussed in this paper. Supporting work such as research on literature and historical sources used GPT 6.1 sol and Gemini 3.8 flash. The authors are responsible for the core judgments, the cases and the final text of the paper.

\bibliographystyle{plainnat}
\bibliography{references}
\end{document}